%% file: Hartin_TIPP09_paper_arxiv.tex
\documentclass[5p]{elsarticle}

\usepackage{amssymb}
\usepackage{graphicx}
\input{mydefs}
\usepackage{booktabs}

\begin{document}

\begin{frontmatter}
\title{Precision Polarimetry at the ILC: Concepts, Simulations and experiments} 
\author[FIRST]{Christoph Bartels, Anthony Hartin, Christian Helebrant, Daniela K\"afer,
Jenny List}
\address[FIRST]{DESY, Hamburg, Germany}

\begin{abstract}

The precision physics program of the ILC requires precise knowledge of the state of beam polarisation.
In fact the Compton polarimeters intended for the ILC will have to measure the polarisation with error a factor 
of 2 smaller than the previous best measurement at the SLAC SLD experiment. In order to 
further reduce measurement error, spin tracking simulations in the ILC Beam Delivery System 
subject to ground motion induced misalignment have been performed and the expected variation in 
polarisation has been quantified. A prototype of a high precision spectrometer to record Compton 
scattered electrons from the interaction of a longitudinal laser and the charged beams has been developed. 
The Compton electrons interact with a gas in the polarimeter channels to produce Cherenkov radiation 
measured by photodetectors. The calibration of the photodetectors is crucial and exhaustive bench tests of 
the photodetector linearity have been performed. The polarimeter prototype itself will be tested at the 
ELSA testbeam in Bonn in Spring 2009.
\end{abstract}


\end{frontmatter}

\section{Introduction}

At the International Linear Collider, polarised electrons and positron beams are foreseen to collide at 
centre-of-mass energies between 90 and 500~GeV. Its physics program aims for extremely precise determination 
of the parameters of the Standard Model of particle physics as well as of yet unknown phenomena. 
This requires very precise beam parameter measurements, such as the beam energy and polarisation, to a 
precision in the order of $10^{-4}$.

While such precision has already been reached for the beam energy measurement at LEP~\cite{LEPIIenergy}, 
the previous best polarisation measurement, performed at the SLAC SLD experiment, had a systematic 
uncertainty of 0.5\%~\cite{SLD}. The ILC, by comparison, will employ a large range of beam energies,
higher intensities and repetition rates. A gain in precision of a factor of 50 in the more demanding 
experimental conditions at the ILC is required. The planned system of upstream and downstream polarimeters 
combined with complementary physics measurements will give an overall precision gain of a factor of 50. 
An upstream polarimeter placed in a purpose designed chicane near the beginning of the Beam Delivery System (BDS)

\begin{figure}[htb]
\centering
\includegraphics[width=0.9\linewidth]{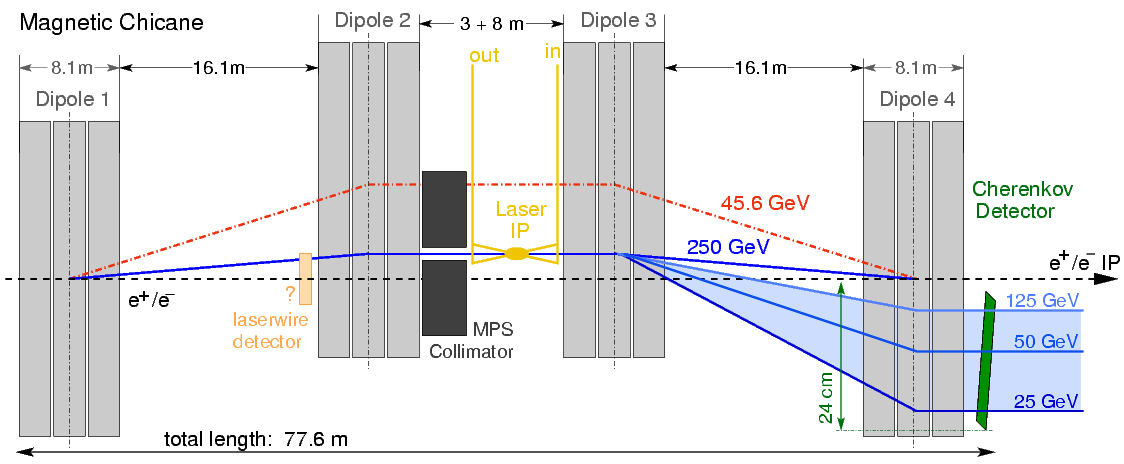}
\caption{Placement of polarimeter in polarimeter chicane.}
\end{figure}

\section{Depolarisation simulations}
In order to combine the measurements of the two Compton polarimeters with each other and with collision data, the depolarisation between these two measurement locations has to be studied. The main depolarisation source is at the Interaction Point (IP) and is due to strong beam-beam effects \cite{ilcdepol}. The Beam Delivery System (BDS) can also induce significant loss of polarisation 
due to ground motion-induced misalignments of its lattice elements. To quantify this latter source of polarisation loss, 
a realistic simulation of depolarisation and spin precession in the BDS is presented.

A realistic bunch train is generated by using PLACET \cite{placet} to track bunches through a misaligned linac in 
which a 1:1 correction and dispersion free steering is made. BDS elements are misaligned using a model of the expected 
inter-train ground motion and the spin is tracked using the BMAD \cite{bmad} program. Finally the expected luminosity-weighted 
depolarisation due to BDS misalignments and the evolution of bunch-to-bunch depolarisation along a bunch train 
corrected by intra-train feedback is investigated. 

\begin{figure}[Htb]
\centering
\includegraphics[width=0.95\linewidth]{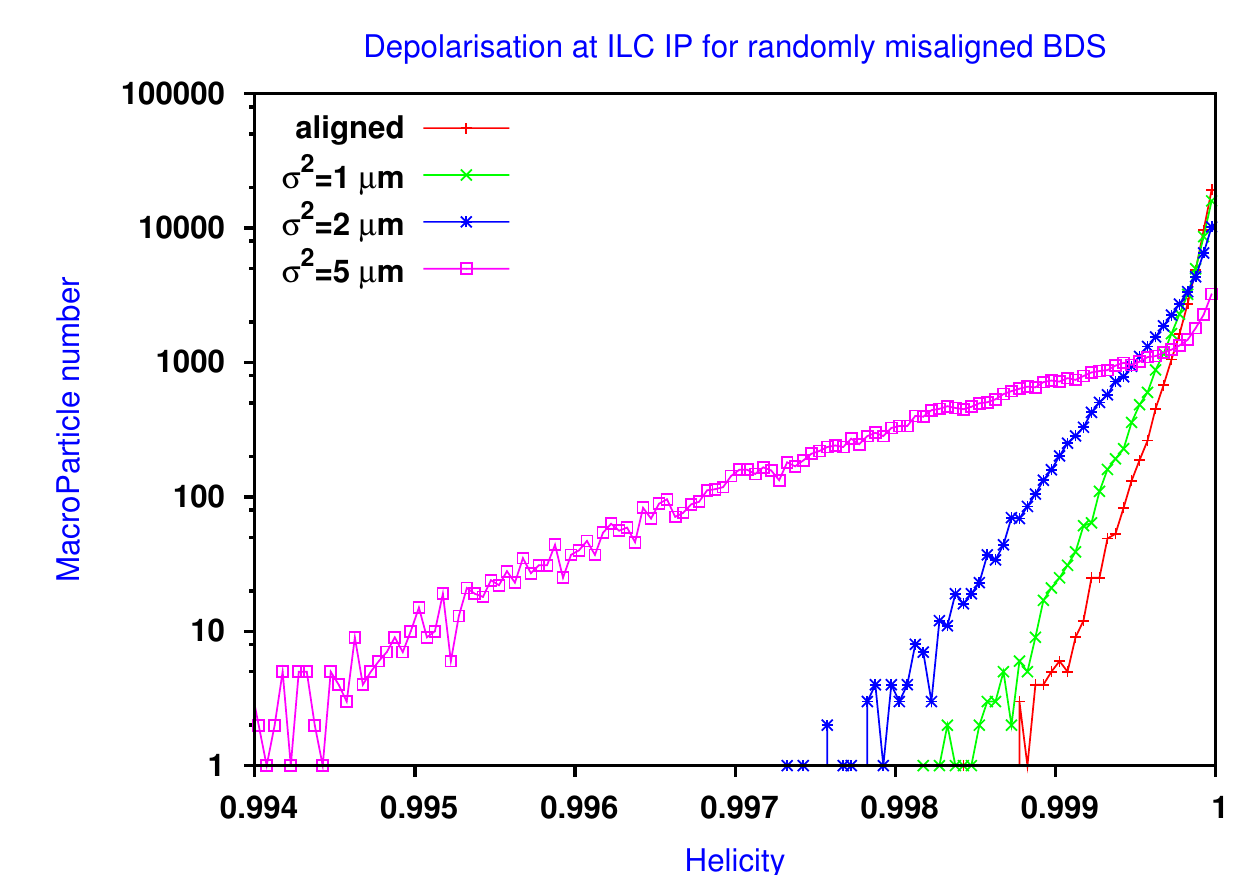}
\caption{\label{depol}Depolarisation at the ILC for a misaligned Beam Delivery System}
\end{figure}

For random misalignments of BDS elements with a variance of 5 microns from true alignment,
the mean helicity of the beam already declines by 0.11\%, with an increasing helicity distribution width (figure \ref{depol}). 
Depolarisation at the ILC is expected to add a further 0.14\% luminosity
weighted depolarisation to the physics collisions. The combined effect of both BDS and IP 
depolarisation makes it necessary to deploy also a downstream polarimeter to provide independent 
information of the polarisation state in physics collisions.

\section{Detector Prototype Design}

The polarimeters measure the polarisation dependent energy spectrum of beam electrons undergoing
Compton scattering with an oncoming laser. Approximately $10^3$ scattered electrons per bunch are produced with 
scattering angles within 10~$\mu$rad w.r.t. to the original beam direction and a magnetic spectrometer 
is employed to transfer the energy distribution into a position distribution.
Behind the spectrometer the electrons are then detected by an array of Cherenkov counters. Since 
knowledge of the analysing power of the whole spectrometer and the linearity of the detector setup were the main 
limiting factors for SLD polarimetry, the design of the Cherenkov detector is a crucial issue for improving 
the precision.  

\begin{figure}[Htb]
\centering
\includegraphics[width=0.95\linewidth]{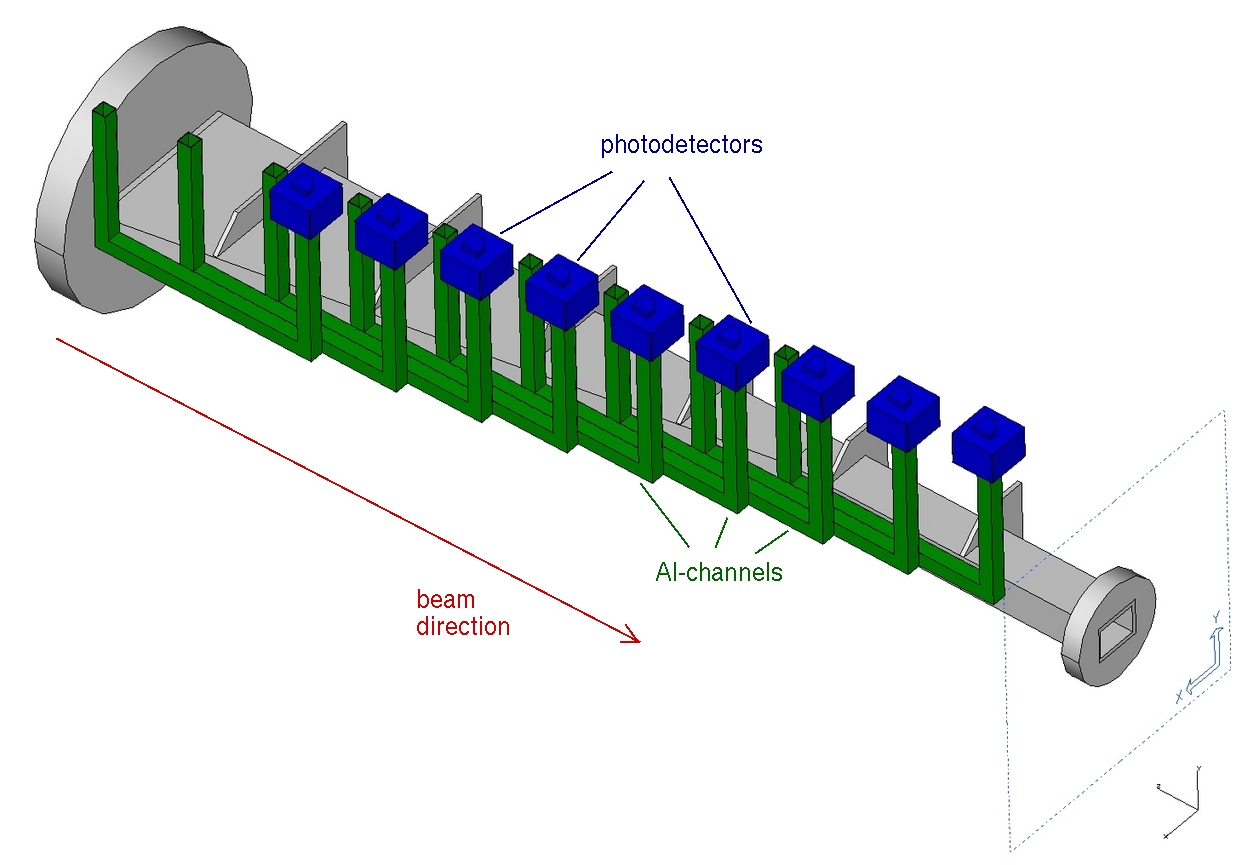}
\caption{New Cherenkov detector design}
\end{figure}

\begin{figure}[Htb]
\centering
\includegraphics[width=0.8\linewidth]{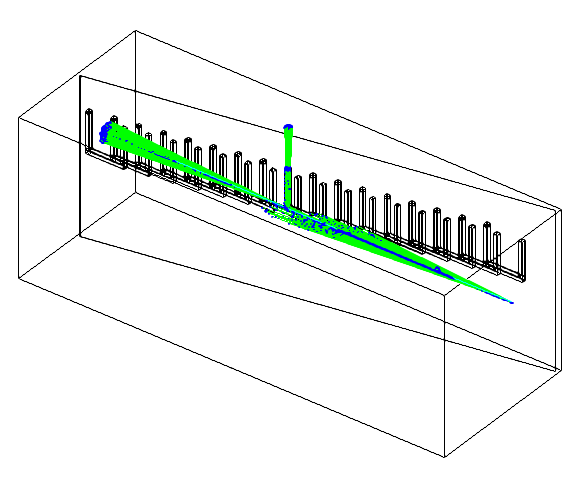}
\caption{\label{ilcproto}Geant simulation of new Cherenkov detector design}
\end{figure}

The Cherenkov detector is planned to consist of staggered, U-shaped aluminium channels along the 
$z$-axis to allow for a tapered beam pipe preventing wake field creation. The channels will have a 
cross section of about 1$\;$cm$\,\times\,$1$\;$cm and are filled with C$_{\rm 4}$F$_{\rm 10}$ as Cherenkov gas. 
One leg of the U-shaped gas tubes is equipped with a photodetector and subsequent readout, while the 
other leg is used for calibration purposes(via LED, or laser light).  Geant simulations have been performed for the planned design 
showing a significant reduction in cross-talk across channels when the detector channels are rotated out of the expected plane of beam synchrotron radiation (figure \ref{ilcproto}).

To study the performance of the proposed Cherenkov detector design, a two channel prototype
was constructed during summer 2008. This was then first tested in the laboratory (using LED,
or laser light) and in a second step in the DESY$\,$II testbeam. Finally, the prototype
was transported to Bonn in spring 2009 and set up at the ``Elektron Stretcher Anlage''
(ELSA). The ELSA testbeam provides higher bunch rates (approximately 7 orders of magnitude higher than for the tertiary DESY$\,$II beam) and the bunches themselves can contain about 100-1000 electrons, a situation closer to that of the linear collider Compton polarimeter design which foresees about 1000 Compton electrons ejected per bunch. The results of these testbeam runs will be reported on elsewhere

\begin{figure}[Htb]
\centering
\includegraphics[width=0.9\linewidth]{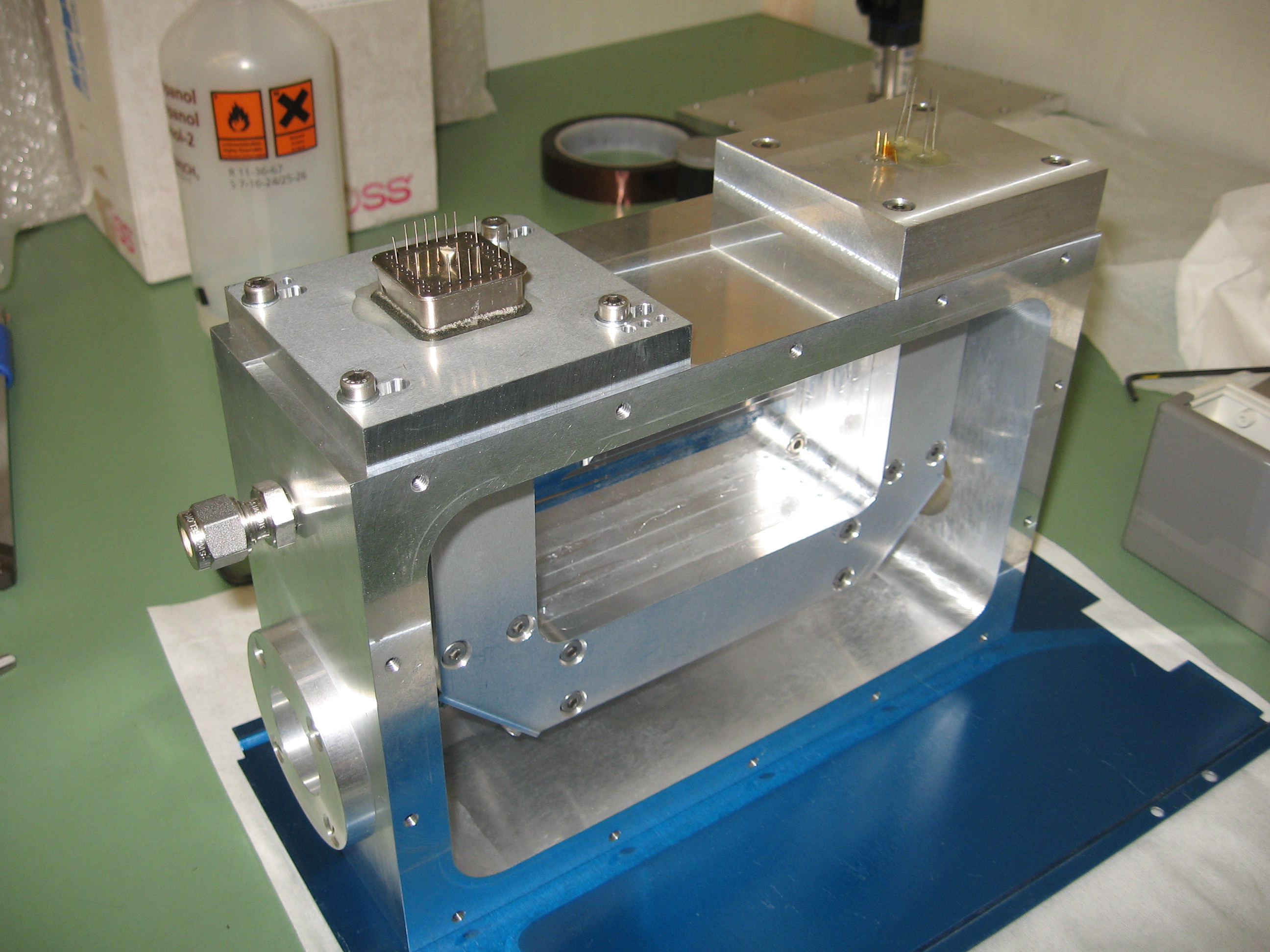}
\caption{Two channel prototype Cherenkov detector}
\end{figure}

\section{Photodetector Studies}

Experiences from previous polarimeters show that the limiting factor of the Cherenkov design will not
be of statistical but systematic nature. The linearity of the entire Cherenkov detector, 
especially the photodetector (PD) is of utmost importance~\cite{bib:phd-elia, bib:nonlin-effects}.

A test facility was set up to analyse different types of PDs regarding their adequacy for
an ILC polarimeter. It consists of a light-tight box, that can be equipped with various types
of PDs ranging from different conventional PD tubes to novel silicon based photomultipliers (SiPM).
Due to the compactness of the latter (some mm$^2$) a much higher spatial resolution of the
Cherenkov detector could be achieved compared to more conventional PDs, and thus a more precise
polarisation measurement. The light is generated by a blue LED connected to a function generator.
The data acquisition is done via VME electronics using a high resolution 12-bit QDC.

\begin{figure}[Htb]
\centering
\includegraphics[width=1.0\linewidth]{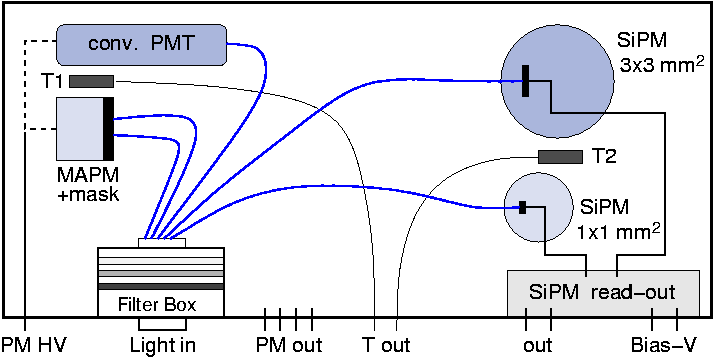}
\caption{Schematic of photodetector test equipment}
\end{figure}

Several methods are employed to measure the integral non-linearity of the PDs. First, a simple array
of calibrated optical filters is used. A second series of measurements is done by varying the
length of an rectangular LED pulse. Two more elaborate methods are exploited to measure the
differential non-linearity of the PDs. For different LED pulse heights $P_i$ and a fixed pulse
$p \ll P_i$ the PD response to $P_i$ and $P_i+p$ is measured. Another approach to measure the
differential non-linearity uses a four-holed mask applied to the PD. An LED pulse is equally
fed into four optical fibres, which can be applied to the four holes in the mask. The DNL is
measured as the difference between the sum of the PD's responses to using only one fibre at a
time and all four together.

\begin{figure}[Htb]
\centering
\includegraphics[width=1.0\linewidth]{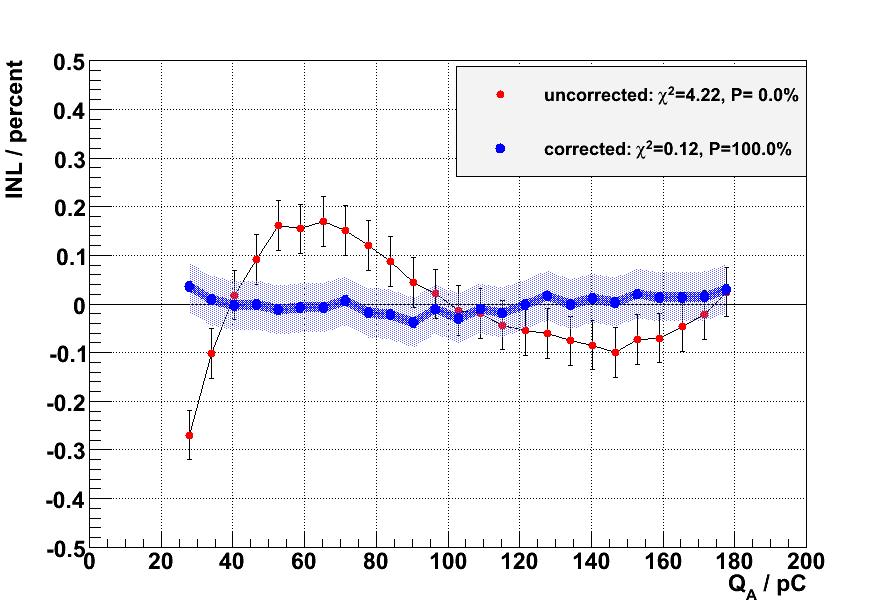}
\caption{\label{inl}Measured PD Integrated non-linearity and applied correction.}
\end{figure}

Once the PD non-linearity has been accurately measured a correction can be applied in order to restore linearity 
on a per-mille level (figure \ref{inl})

\section{Conclusion}

Spin tracking simulations show a significant depolarisation in the beam delivery system of a planned future linear collider and so both an upstream and downstream polarimeter will be necessary to establish the polarisation state of particle collisions at the IP of a linear collider. Additionally, the required physics measurements will neccesitate a polarimeter Cherenkov detector of unprecedented accuracy. A new detector prototype has been designed with the detector arms formed into a staggered U shapes. Geant simulations have shown a reduction in channel cross-talk compared to previous designs. The accuracy requirements dictate that the non-linearity of the photodetectors used in the Cherenkov detector be well understood and correctable. A laboratory testbox was developed and correction of the measured photodetector non-linearity was achieved to a per-mille level. A 2 channel detector prototype was setup and tested in the DESY testbeam and the ELSA facility in Bonn.

\section*{Acknowledgements} This work has been supported by the DFG via the Emmy-Noether grant Li 1560/1-1.

\begin{footnotesize}


\end{footnotesize}


\end{document}

%% file: mydefs.tex
\newcommand{\nwcd}{\newcommand}
\nwcd{\dfrac}{\frac}
\nwcd{\tfrac}{\frac}
\nwcd{\dint}{{\displaystyle \int}}
\nwcd{\dsum}{\sum}
\nwcd{\text}{\mbox}
\nwcd{\limfunc}{\mbox}
\nwcd{\NEG}{\slash}
\nwcd{\nts}{\negthinspace}
\nwcd{\nms}{\negmedspace}
\nwcd{\nths}{\negthickspace\negthickspace}
\nwcd{\vep}{\varepsilon}
\nwcd{\undl}{\underline}
\nwcd{\dd}{\ddagger}
\nwcd{\pr}{\prime}
\nwcd{\witl}{\widetilde}
\nwcd{\bm}{\pmb}
\nwcd{\mvb}{\mathversion{bold}}
\nwcd{\lami}{(\lambda_{i})}
\nwcd{\lamf}{(\lambda_{f})}
\renewcommand{\bar}[1]{\setbox0=\hbox{$#1$}
                       \setlength{\unitlength}{\wd0}
                       \smash{\buildrel \line(1,0){0.7} \over #1}}
\def\dbr#1{\setbox0=\hbox{$#1$} 
           \setlength{\unitlength}{\wd0}
           \smash{\buildrel \overline{\line(1,0){0.7}} \over #1}}
\nwcd{\balph}[1]{\setbox0=\hbox{$#1$}
                       \setlength{\unitlength}{\wd0}
                       \smash{\mathrel{\mathop{\line(1,0){0.7}} \atop #1}}}

\nwcd{\half}{{\textstyle \frac{1}{2}}}
\nwcd{\sql}[1]{#1\hspace{-1.5ex}\raisebox{-1.0ex}{$\sim$}}
\nwcd{\squ}[1]{#1\hspace{-1.5ex}\raisebox{-1.4ex}{$\sim$}}
\nwcd{\eapbf}{\frac{e^{2}a^{2}}{4(k\bar{p})(kp_{f})}}
\nwcd{\eapbi}{\frac{e^{2}a^{2}}{4(k\bar{p})(kp_{i})}}
\nwcd{\eapdbf}{\frac{e^{2}a^{2}}{4(k\dbr{p})(kp_{f})}}
\nwcd{\eapdbi}{\frac{e^{2}a^{2}}{4(k\dbr{p})(kp_{i})}}
\nwcd{\pipbrp}{[(p_{i}\bar{p}_{r\pr})+m^{2}]}
\nwcd{\pipbr}{[(p_{i}\bar{p}_{r})+m^{2}]}
\nwcd{\bpr}{\bar{p}_{r}}
\nwcd{\dpr}{\dbr{p}_{r\pr}}
\nwcd{\pmsqb}{\bar{p}^{2}_{r}-m^{2}}
\nwcd{\pmsqbp}{\bar{p}^{2}_{r\prime}-m^{2}}
\nwcd{\pmsqdb}{\dbr{p}^{2}_{r}-m^{2}}
\nwcd{\pmsqdbp}{\dbr{p}^{2}_{r\prime}-m^{2}}
\nwcd{\mat}[1]{$\; #1 \;$}
\nwcd{\gmu}{\gamma_{\mu}}
\nwcd{\gnu}{\gamma_{\nu}}
\nwcd{\gmaok}{\gamma_{\mu}\slashed{a}_{1}\slashed{k}}
\nwcd{\gmatk}{\gamma_{\mu}\slashed{a}_{2}\slashed{k}}
\nwcd{\gnaok}{\gamma_{\nu}\slashed{a}_{1}\slashed{k}}
\nwcd{\gnatk}{\gamma_{\nu}\slashed{a}_{2}\slashed{k}}
\nwcd{\aokgm}{\slashed{a}_{1}\slashed{k}\gamma_{\mu}}
\nwcd{\atkgm}{\slashed{a}_{2}\slashed{k}\gamma_{\mu}}
\nwcd{\aokgn}{\slashed{a}_{1}\slashed{k}\gamma_{\nu}}
\nwcd{\atkgn}{\slashed{a}_{2}\slashed{k}\gamma_{\nu}}
\nwcd{\pmsq}{\displaystyle \left[\frac{\slashed{p} + m}{p^{2} - m^{2}}
\right]}
\nwcd{\pbr}{\left[\frac{\slashed{\bar{p}}_{r} + m}{\bar{p}_{r}^{2} - m^{2}}
\right]}
\nwcd{\pbrp}{\left[\frac{\slashed{\bar{p}}_{r\prime} + m}{\bar{p}_{r\prime}
^{2} - m^{2}}\right]}
\nwcd{\pdbr}{\left[\frac{\slashed{\dbr{p}}_{r} + m}{\dbr{p}_{r}^{2} - m^{2}}
\right]}
\nwcd{\pdbrp}{\left[\frac{\slashed{\dbr{p}}_{r\prime} + m}{\dbr{p}_{r\prime}
^{2} - m^{2}}\right]}